\documentclass[prx,aps,twocolumn,superscriptaddress]{revtex4-2}

\usepackage[dvips]{graphicx}
\usepackage{amssymb,amsfonts,amsmath,gensymb,eucal,bm}
\usepackage{color}
\usepackage[normalem]{ulem}
\usepackage{natbib}
\usepackage[hidelinks]{hyperref}
\usepackage[capitalize]{cleveref}
\crefformat{equation}{equation~(#2#1#3)}

\renewcommand{\vec}[1]{\boldsymbol{#1}}

\newcommand{\ind}[1]{\ensuremath_{\mathrm{#1}}}
\renewcommand{\sup}[1]{\ensuremath^{\mathrm{#1}}}

\newcommand{\vy}{\ensuremath{{\bf y}}}
\newcommand{\vx}{\ensuremath{{\bf x}}}
\newcommand{\vb}{\ensuremath{{\bf b}}}
\newcommand{\vyb}{\ensuremath{\vec{\overline{y}}}}
\newcommand{\vyh}{\ensuremath{\vec{\hat{y}}}}
\newcommand{\vxb}{\ensuremath{\vec{\overline{x}}}}
\newcommand{\norm}[1]{\ensuremath{\left\Vert #1 \right\Vert}}

\newcommand{\athing}[2]{\ensuremath{\underset{#1}{\mathrm{arg\,#2}}}}
\newcommand{\amin}[1]{\athing{#1}{min}}
\newcommand{\amax}[1]{\athing{#1}{max}}
\newcommand{\mat}[1]{\ensuremath{\bm{#1}}}

\DeclareMathOperator{\sgn}{sgn}

\newcommand{\affil}[2]{#1, Utrecht University, Princetonplein #2, 3584 CC Utrecht, The Netherlands}

\newcommand{\SCMB}{\affil{Soft Condensed Matter and Biophysics, Debye Institute for Nanomaterials Science}{1}}

\graphicspath{{figures}}
\begin{document}

\title{Physical Neural Networks Need Nonlinearity, Amplification, and Suppression for Learning}

\author{Nex C.\ X.\ Stuhlm\"uller}\email{n.c.x.a.n.stuhlmuller@uu.nl} \affiliation{\SCMB{}}
\author{Marjolein Dijkstra} \email{m.dijkstra@uu.nl}\affiliation{\SCMB{}}

\begin{abstract}
The exponential growth in energy consumption of artificial intelligence systems has spurred interest in physical computing paradigms that exploit the relaxation of physical systems toward steady states. However, many existing physical networks are fundamentally linear and incapable of performing nonlinear operations crucial for meaningful machine learning tasks. Here we use simulations to show that nonlinearity alone is insufficient; physical learning systems must also support signal amplification and suppression to perform nontrivial computations. We present physically plausible circuit designs that incorporate  these essential features, enabling effective nonlinear information processing. Our findings clarify the limitations of linear physical networks and provide guidance for developing energy-efficient physical learning architectures capable of general machine learning tasks.
\end{abstract}

\maketitle

\section{Introduction}
In 2012, Alex Krizhevsky {\em et al.}\ reignited interest in artificial intelligence (AI) and brought spring to the AI-winter with the release of AlexNet, which outperformed all other competitors in the Image Large Scale Visual Recognition Challenge (ILSVR) by more than $10$ percentage points~\cite{Krizhevsky2017,Lab2010}. The ILSVR, held  annually from  2010 to 2017,  challenged participants to classify more than one million images  into 1000 categories. AlexNet's  outstanding performance was achieved by the use of GPUs for training, providing substantially more computational power than competing classifiers. Since then, the energy consumption of AI systems has doubled roughly every four  months~\cite{Tripp2024}, with a further acceleration since the massive scaling of large language model operations starting in 2023.
 If this trend of doubling approximately every four  months continues, the energy demands of AI systems would exceed the total global energy production within the next decade.

To combat this unsustainable growth, researchers have increasingly drawn inspiration from the biological brain, giving rise to the field of neuromorphic computing, in which machine learning is implemented using systems that more closely mimic the information processing mechanisms of the brain~\cite{Markovic2020}. Examples include  analog computing architectures~\cite{Zhang2026}, spiking neural networks~\cite{Yamazaki2022}, and ionic rather than  electronic information carriers~\cite{Chun2015}.  
One emerging subfield is physical computing, in which the relaxation of physical systems toward steady states is  exploited to perform machine learning tasks.
Although the field remains largely in its infancy, several  impressive advances have been reported recently~\cite{Du2026,Shohat2025,Doremaele2023,Bos2026,Taglietti2026,Zolfagharinejad2025}.
One particularly intriguing example  is the classification of the Iris dataset using a linear network trained solely through local learning rules, thereby circumventing the energy-intensive  backpropagation algorithm~\cite{Nachi}.
Since nontrivial classification is inherently a nonlinear task, we investigate here how  linear networks can  perform successful classification~\cite{Anisetti2023,Nachi}.

After a brief introduction to machine learning (ML), we show how these networks perform linear tasks, identify the mechanisms that make  linear networks appear to perform classification, and finally clarify what is  required for physical networks to successfully carry out machine learning tasks.
\section{Machine learning basics}
\label{sec:mlintro}
Neural networks are a tool for approximating a desired function $f_d({\bf x})$ of some input ${\bf x}$ by a parametrized model  $f_n({\bf x}|P, \{f_a\})$, where the   parameters are given by $P=\left\{\mat{W}^i | i \in [1,N]\right\}$ and $\{f_a\}$ denotes the set of activation functions.
The network topology and the choice of activation functions $\{f_a\}$  determine the class of functions that the network can approximate.

The universal approximation theorem in machine learning states that, given a non-polynomial activation function (e.g.\ ReLu or sigmoid), any $L^p$ function can be approximated to arbitrary precision by a sufficiently large neural network \cite{Cai2022}. Here, an $L^p$ function denotes a function $f$ for which the $p$-th power of the absolute value is integrable, i.e. $\int |f(x)|^p dx$ is finite.

\begin{figure}[htbp]
  \centering
  \includegraphics[width=0.8\linewidth]{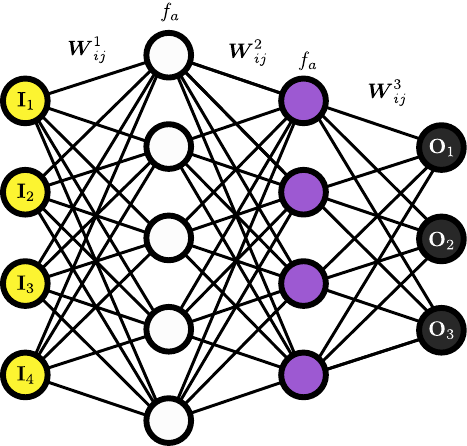}
  \caption{Machine learning setup used in this work. The 
  yellow nodes represent the input layer $I_i,~i\in\{1\dots 4\}$. The input signals are weighted by the matrices $W_{ij}^1$ and passed to the first hidden layer  (white circles), where the activation function $f_a$ is applied. The resulting outputs are then transmitted to the second hidden layer (purple circles), where the activation function $f_a$ is applied. Finally, the outputs of $f_a$ are weighted by  $W_{ij}^3$ and  interpreted as the output nodes (black circles) $O_i,~i\in\{1\dots 3\}$.}
  \label{fig:ml-setup}
\end{figure}

In principle, a different activation function can be chosen for each node and layer.
For simplicity, we use the same activation function $f_a$ for every node in the network:
\begin{equation}
  \label{eq:mlchain}
f_d(\vx) \approx f_n(\vx)=\mat{W}^N\cdot f_a.\left(\mat{W}^{N-1} \cdots f_a.(\mat{W}^1 \vx)\right),
\end{equation}
where the dot $.$ between the function name and its argument denotes element-wise application.
In this work, we use networks with two hidden layers containing  5 and 4 nodes, respectively, as illustrated in \cref{fig:ml-setup}. The number of input nodes and  output nodes are chosen to match the dimensions of the corresponding dataset.

The activation functions must either be non-polynomial or polynomial with leading exponent greater than one.
If the activation functions $f_a$ are chosen to be linear or affine-linear, the entire sequence of operations collapses into a single matrix-vector multiplication, reducing the network to a linear transformation
\begin{equation}
  \label{eq:linearml}
  f_n(\vx)=\prod_{j}^N \mat{W}^j\vx = \mat{W}\ind{comb}\vx,
\end{equation}
where $\mat{W}\ind{comb}=\prod_{j=1}^N \mat{W}^j$ is the product of all weight matrices.
Note that we omit explicit bias vectors here, since they can be incorporated trivially by appending a constant 1 to the input vector ${\bf x}$ and extending the weight matrices $\mat W$ accordingly,   $ \left( \mat W || \vb \right) \cdot \begin{pmatrix} \vx \\ 1 \end{pmatrix} = \mat W \cdot \vx + \vb$.
As we will show later, many physical machine-learning networks do not employ activation functions and therefore reduce to purely linear networks, making them fundamentally incapable of performing nonlinear tasks.

A training dataset $S\sup{tr} = \{{\bf y}\sup{tr},{\bf x}\sup{tr}\}$ is used to optimize the parameters $P$.
During training, a sample $S^i = ({\bf x}^i, {\bf y}^i)$ is randomly selected from $S\sup{tr}$,  and the input ${\bf x}^i$ is fed into the network, producing the output $\hat{{\bf y}}^i= f_n({\bf x}^i)$. This output is then compared with the desired output ${\bf y}^i$ using a loss function $L({\bf y}^i, \hat{{\bf y}}^i)$. Subsequently, the network parameters $P$ are adjusted to reduce the loss.
Typically, this optimization is performed using gradient descent on the loss function, iterating until either a predefined number of training epochs has been performed or the performance on the training dataset reaches a desired threshold.

Next, we briefly discuss how linear physical networks that lack intrinsic  nonlinearities can nevertheless appear to perform  nonlinear tasks. A more detailed treatment, including exact update procedures, is provided in Appendix \ref{sec:contrastive}.

\section{How linear physical networks appear to perform nonlinear tasks}
Flow networks composed of linear elements, such as Ohmic resistors in electrical circuits, Hagen–Poiseuille channels in fluidic systems, or mechanical mass-spring networks, are themselves linear systems as a whole.
Consequently, their outputs cannot be more complex than  linear combinations of their inputs, and the entire network can be represented by \cref{eq:linearml}, or more compactly as 
\begin{equation}
    \vy = \mat{W}\vx .
\end{equation}
Such networks can therefore perform simple linear tasks, such as  fitting one straight line. Here,  we show that their performance deteriorates significantly for multivariate inputs (see Appendix \ref{sec:contrastive} and \cref{fig:linear}).
This naturally raises the question of how such networks appear to learn in the first place.

One approach to physical learning that has gained significant attention  in recent years is contrastive learning, in which two states of a network, a free state and a clamped state, are compared, and the weights are updated according to a purely local learning rule.
It has been shown that this local update rule approximates gradient descent~\cite{Stern2021}.
This approach offers two major advantages. First, it avoids the need to explicitly perform backpropagation, thereby reducing one of the dominant energy costs associated with training neural networks. Second, it avoids the difficulty that the behavior of physical systems is often not known in a closed mathematical form, making it impossible (or impractical) to compute exact derivatives of the system output with respect to its parameters. Such derivatives are required for conventional gradient-based optimization methods like backpropagation.
We will explain this local update rule in detail in Appendix  \ref{sec:contrastive}; for now, the precise details are not essential.
One important practical aspect, however, is that each network weight can only change in finite increments, since real-world implementations often use digital potentiometers (i.e. scalable resistors) with discrete resistance values.
In this work, we consider resistors with either 128 (as in Ref.~\cite{Nachi}) or 1280 evenly spaced resistance values, in order to test whether the performance of the network is constrained by the finite resolution of the resistive elements or by more fundamental factors.

\subsection{Linear tasks}
\begin{figure}
    \centering
    \includegraphics[width=\linewidth]{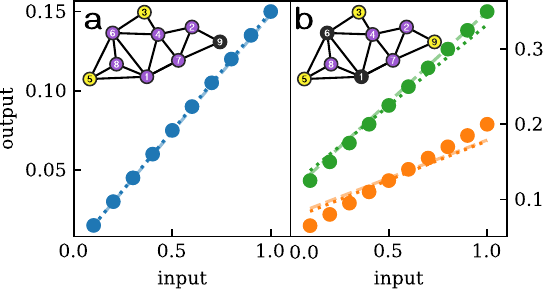}
    \caption{Linear network performing linear regression. Insets in the top show the network used for training. Yellow nodes are input nodes, black nodes are output nodes, and purple nodes are hidden nodes. Solid dots in (a,b) denote the true labels, while dashed lines indicate the predicted function using a resistor resolution of 128 levels per element, and dotted lines show the prediction using 1280 levels per element. (a) A single linear function (blue) is learned successfully, achieving high accuracy ($\mathrm{MSE}\approx 10^{-5}$), which further improves when increasing the resolution of resistors ($\mathrm{MSE}\approx 10^{-7}$). (b) Two linear functions  (orange and green) are learned simultaneously. The accuracy is worse than in (a) ($ \mathrm{MSE}\approx 10^{-3}$) and does not improve with higher resistor resolution.
    The insets show the network architecture used for learning. Yellow nodes correspond to input nodes, purple nodes to hidden nodes, and black nodes to output nodes.}
    \label{fig:linear}
\end{figure}

We first show that physical linear networks can, to some extent, perform linear tasks.
We use a physical network similar to Ref.~\cite{Nachi} (shown in the inset of \cref{fig:linear}) and consider two different tasks.
In the first task,  we apply an input signal (node 3) measured relative to a constant reference signal (node 5), and read out the network output  at node 9.
The network is trained using the contrastive learning rule (i.e.\ \cref{eq:main}) to realize the linear mapping $O(I)=0.15I$, where $I$ denotes the input signal  applied to node 3.
The network performs this task accurately, achieving a mean squared error $\mathrm{MSE}\approx 10^{-5}$, and its performance further improves as the resistor resolution is increased (see \cref{fig:linear} (a)).
In the second task,  we train the network to produce two outputs $O_1$ and $O_2$, corresponding to nodes 1 and 6, respectively, as a function of two inputs $I_1$ and $I_2$ (nodes 3 and 9, respectively). The target mappings are $O_1=0.15I_1 + 0.1I_2$ and  $O_2=0.2I_1 + 0.25I_2$, respectively. In this case, the overall performance is mediocre. The network achieves only an  $\mathrm{MSE}\approx 10^{-3}$, which corresponds to significant deviations from the target linear functions. This behavior is consistent with the work of Ref.~\cite{Nachi}.
To show that the limiting factor is not the finite resolution of the resistors but the learning approach itself, we increased the resistor resolution by an order of magnitude and retrained the network. However,  the network error did not reduce (compare the dotted and dashed lines in~\cref{fig:linear}). We therefore conclude that these networks perform poorly even on tasks for which linear architectures should, in  principle, be a good choice.
Using two separate networks for the two outputs would likely work significantly better, since it would eliminate crosstalk between the outputs.

\subsection{Classification}
A classifier typically assigns a numerical score to each possible class label for a given sample. The class associated  with the highest score is typically interpreted as the predicted label.
To classify a sample correctly, the signal corresponding to the assigned class must therefore be distinguishable from all competing class signals. In practice, this requires the network to produce a comparatively large output for the correct class while simultaneously suppressing the outputs associated with all other classes.

This intrinsic requirement of selectively enhancing one signal relative to the others makes classification fundamentally incompatible with architectures that cannot amplify or suppress signals~\cite{Chorowski2015}. In purely passive physical systems, outputs cannot exceed the magnitude of the inputs (no amplification), and signals generally cannot be subtracted from one another (no suppression). As a consequence, such systems lack the basic mechanisms required for robust classification (see also Appendix \ref{sec:approx}).
To circumvent these limitations, one may instead employ  a  nonlinear readout procedure based on the Euclidean ($L_2$) distance between a sample output and the mean output associated with each class, as employed in Ref.~\cite{Nachi}. Because this procedure relies on the 2-norm, we refer to it here as the $L_2$-readout. Importantly, this readout must be performed externally on a conventional computer and requires additional memory to store the class averages.
The $L_2$-readout proceeds as follows 
\begin{enumerate}
    \item Feed all training samples belonging to class $j$ through the network.
    \item Record the corresponding network outputs and compute their  average, yielding the average output $\vec{\overline{y}}_j$ for class $j$.
    \item Repeat this procedure for all classes.
    \item To classify a new sample $\vx_i$, feed it through the network to obtain the output  $\hat{\vy}_i$. Then compute the Euclidean distances $||\hat{\vy}_i-\overline{\vy}_j||$ for each class $j$.
    \item Assign the sample the predicted label $l_p$ corresponding to the class $j$ yielding the minimal Euclidean distance. 
\end{enumerate}
Formally, this procedure can be written as 
\begin{equation}
l_p = \amin{j} \norm{\vyh_i-\vyb_j} = \amin{j} \norm{\mat{W}( \vx_i - \vxb_j)},
\end{equation}
where $\vxb_j$ denotes the average of all samples belonging to class $j$, and $\vyb_j = \mat{W}\vxb_j$.
This expression already suggests  that the inclusion of the linear map $\mat{W}$, and therefore the underlying flow network, will only rarely change the predicted class as compared to no linear map/flow network being present. A linear transformation can only rotate, scale, or  shear its input space. Rotations and uniform scalings preserve the relative ordering of distances between samples and class means, and therefore do not affect the classification result. Consequently, the only potentially relevant transformation induced by $\mat{W}$ is a shear.  
However, for typical high-dimensional statistical datasets, samples from each class are distributed approximately isotropically around their respective means. In other words, the spread of each class is roughly uniform in all directions, rather than being strongly elongated along particular axes. Under such conditions, even if $\mat{W}$ shears different directions unequally, the relative magnitudes of the transformed distances $\mat{W}(\vx_i-\vxb_j)$ remain largely unaffected. Below, we will show that the classifier performance arises primarily from the external nonlinear $L_2$-readout rather than from the linear physical network itself.

To evaluate the contribution of a network that transforms the input data before applying the readout, its performance should be compared to the baseline case in which no network is used and the raw input data is  passed directly to the $L_2$-readout. In the formalism above, this corresponds to choosing the transformation matrix $\mat{W}$ to be the identity matrix $\mathbb{I}$. 
To efficiently optimize $\mat{W}$, we slightly modify the readout procedure. Instead of assigning the predicted label according to $l_p = \amin{j}\left\Vert \vyb_j -\vy\right\Vert$, we use $l_p=\amax{j} \left(\vyb_j \cdot \vy\right)$. For normalized data, these two formulations are equivalent, but the latter  is more convenient for machine-learning optimization methods. Because physical systems can only realize bounded signal amplitudes, we rescale all inputs to lie within the interval (0,1) without loss of generality. We then optimize $\mat{W}$ using gradient descent with logistic cross entropy as the loss function, while leaving $\mat{W}$ unconstrained by physical realizability. This provides an upper bound of the achievable performance of a linear network combined with the $L_2$-readout.
When we instead attempted  to directly optimize the original $L_2$-distance readout by minimizing the distances themselves, all weights collapsed to zero. Intuitively, the optimization lacks a mechanism that consistently rewards increasing separation between classes, leading to vanishing weights. Adding regularization terms did not resolve this issue or produce meaningful learning behavior. For this reason, we adopted the arg max-based readout together with logistic cross-entropy as the training objective.

We perform classification of several basic machine-learning datasets, including Iris (4 features, 3 classes, 120 samples)~\cite{iris_53}, Penguins (4 features, 3 classes, 342 samples)~\cite{Penguins}, Wine (13 features, 3 classes, 178 samples)~\cite{wine_109}, Rice (7 features, 2 classes, 3810 samples)~\cite{rice_545},  Wholesale Customers (7 features, 2 classes, 440 samples)~\cite{wholesale_customers_292}), and Magic (10 features, 2 classes)~\cite{magic_gamma_telescope_159}.
We list the classification accuracies in Table \ref{tab:L2readout}.
Surprisingly, we find that the readout function is already extremely powerful on its own. Even without a neural network transforming the input data, it can classify several basic machine-learning datasets with accuracies exceeding $90\%$, including Iris, Penguins, Wine, Rice, and Wholesale Customers.
When comparing the performance with and without an optimized linear map $\mat{W}$, we observe in \cref{tab:L2readout} that the network improves the classification accuracy by only about $2\%$, even when $\mat{W}$ is allowed to be an arbitrary linear transformation that is unconstrained by  physical realizability.
We expect that imposing realistic physical constraints on  $\mat{W}$  would reduce the performance even further. 
The strong baseline performance of the $L_2$-readout itself therefore appears to be  the primary reason why the network of Ref.~\cite{Nachi} can still appear to train successfully, even when gradient ascent (!) is used or when the actual  network output is ignored during training (see Appendix \ref{sec:contrastive} for details).

We also compare the performance of the linear network with that of a multiclass perceptron (MCP), since the two architectures are mathematically closely related.
A multiclass perceptron assigns a predicted label $l_p$ to an input vector $\vx$ according to
\begin{equation}
    l_p = \amax{j}\left(\vec{w}_j \cdot \vx +b_j\right),
    \label{eq:mcp}
\end{equation}
where ${\bf w}_j$ are trainable weight vectors and the index $j$ labels the corresponding class.
MCPs are equivalent to linear classifiers, which becomes clear by rewriting \cref{eq:mcp} as $l_p = \amax{j}\left(\mat{W} \vx + \vec{b}\right)_j$, where the rows of $\mat{W}$ are given by the vectors  ${\bf w}_j$, and ${\bf b}$ is composed of the bias terms $b_j$ of \cref{eq:mcp}.
Note that both the MCP and the $L_2$-readout procedure possess a similar number of trainable parameters.
For example, in the classification of the Iris dataset, the MCP has 15  free parameters ([4 inputs + 1 bias] $\times$ 3 classes), which is comparable to the 16 free parameters (i.e. tunable resistors) of the linear network.

\begin{table}[htbp]
  \centering
  \begin{tabular}{l|cccccc}
    Dataset & $L_2$ & $\mat{W} L_2$ & MCP & npML & spML& ML\\
    \hline

 Iris       & $97\%$  & $96\%$  &  $98\%$  & $96\%$  &  $97\%$  & $100\%$\\
 Penguins   & $99\%$  & $99\%$  & $100\%$  & $99\%$  & $100\%$  & $100\%$\\
 Wine       & $99\%$  & $99\%$  &  $98\%$  & $98\%$  &  $99\%$  & $100\%$\\
 Rice       & $92\%$  & $94\%$  &  $93\%$  & $94\%$  &  $93\%$  &  $94\%$\\
 Customer   & $90\%$  & $90\%$  &  $93\%$  & $92\%$  &  $93\%$  &  $94\%$\\
 Magic      & $77\%$  & $80\%$  &  $78\%$  & $87\%$  &  $87\%$  &  $87\%$\\
 \hline
Yin Yang    & $76\%$  & $76\%$  &  $74\%$  & $96\%$  &  $98\%$  &  $97\%$\\
 Circles    & $53\%$  & $61\%$  &  $61\%$  & $99\%$  & $100\%$  &  $99\%$\\
  \end{tabular}
  \caption{Classification accuracies for the different classifiers considered in this work. For each classifier-dataset combination, 25 independent training runs were performed, and the best-performing result is  reported. The $L_2$ column corresponds to classification obtained by directly applying the $L_2$-classifier to the input data without any trainable network. The $\mat{W} L_2$ column includes an additional trainable linear  mapping before the $L_2$-classifier. The MCP column reports the accuracies of trained multiclass perceptron networks. The npML results correspond to physical 
  machine learning with the normalization regularization scheme, whereas spML denotes physical machine learning using  a symmetrized activation function. Finally, ML refers to a conventional machine-learning network.}
  \label{tab:L2readout}
\end{table}

\begin{figure}
  \centering
  \includegraphics[width=\linewidth]{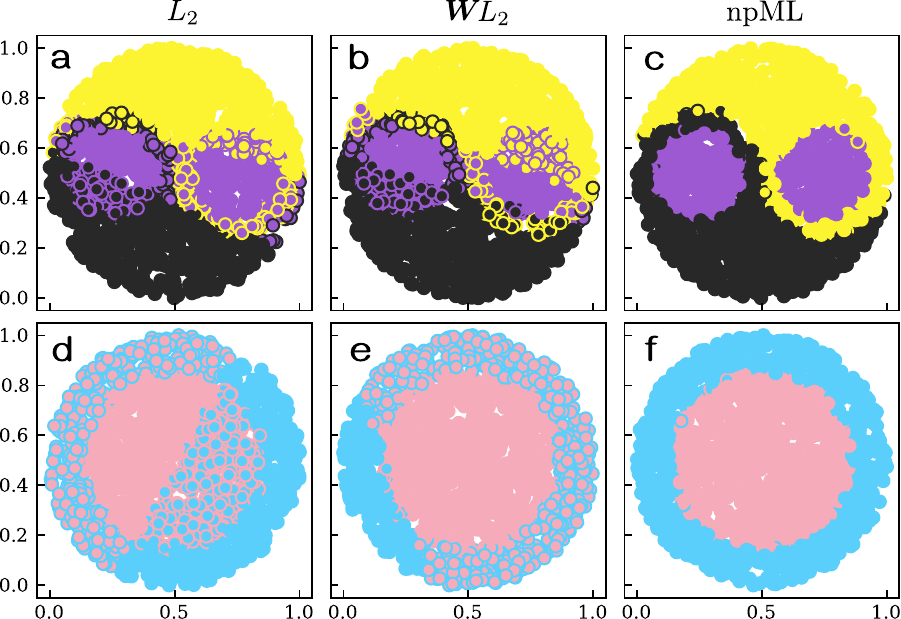}
  \caption{Comparison of three classification approaches  on two datasets. The top row (a,b,c) shows results for the non-binary Yin-Yang dataset, while the bottom row (d,e,f) corresponds to the Circles dataset. The first column (a,d) shows the results of a pure $L_2$-readout applied directly to the input data, without any preceding network. The  second column (b,e) shows a network consisting of a linear transformation followed by the same classifier, i.e.\ $\mat{W} L_2$. The third column (c,f) corresponds to a physically realistic nonlinear ML network.  For each datapoint, the color of the inner circle indicates the predicted label, whereas the color of the outer ring denotes the true label. Points for which the inner and outer colors differ therefore correspond to misclassifications. 
  }
  \label{fig:yinyang}
\end{figure}

To test the performance of the linear network on datasets that are not trivially classified by the $L_2$-readout, i.e.  datasets that are  not linearly separable, we consider the non-binary (ternary) Yin-Yang dataset~\cite{Kriener2021} and the Circles dataset consisting of two classes corresponding to two concentric circles, as  shown in \cref{tab:L2readout} and \cref{fig:yinyang}.
For the Circles dataset, we obtain a classification accuracy of only $61\%$,  which is only marginally better than random guessing. For the Yin-Yang dataset, the accuracy reaches $75\%$, still far below the performance expected from genuinely nonlinear classifiers.
It is tempting to conclude that the linear networks perform well on the 
Yin-Yang dataset, given their  accuracy of $75\%$, which is more than twice the random-guessing baseline. However, this apparent performance is misleading. In \cref{fig:yinyang}, we compare the predicted  and  true label for each datapoint of  the Yin-Yang and Circle dataset. The color of the outer ring denotes the true label, whereas the color of the inner circle indicates the predicted label. Points for which the inner and outer colors differ therefore correspond to misclassifications. We clearly find that the network  does not capture the underlying structure of the dataset at all. Since the network is linear, it can only produce linear decision boundaries and therefore cannot represent the curved class boundaries inherent to the 
Yin-Yang geometry. The reported accuracy thus reflects a coarse projection of the problem rather than a faithful representation of its structure. This highlights the importance of  fairly assessing the quality of networks, rather than relying on  seemingly high classification accuracies~\cite{Anisetti2023}.
In Appendix \ref{sec:systemsize}, we demonstrate that increasing the network size  does not improve the learning, neither on trivial datasets nor on 
nontrivial ones.
We therefore conclude that linear networks are not suitable for most meaningful classification tasks.
This naturally raises the question of how such systems can be modified or extended to enable genuine, nontrivial  machine learning.

\section{Essential Requirements for Physical Learning in  Physical Systems}

As we have seen, purely linear networks are not general-purpose ML devices. This motivates  the introduction of  nonlinearities into the network architecture. To maintain close to conventional machine learning frameworks and to leverage the existing body of knowledge generated in that field,  we construct networks in which the connections (edges) remain linear, while the node operations are nonlinear. In this formulation, when an input potential $x$ is applied to a node, it produces an output $f_a(x)$ that is passed to the next layer, where $f_a$ denotes a nonlinear activation function. To remain consistent with  physical constraints, specifically  that physical potentials are bounded by a maximum achievable potential  in the system, we restrict the range of the activation function to $f(\mathbb{R}) \subset (0,1)$ without loss of generality. Sigmoidal activation functions naturally satisfy this requirement. We therefore choose  $f_a(x)=\sigma(5x)$ as the activation function for our physical networks. Note that sigmoid-like activation functions can also be implemented in hardware settings~\cite{Amin2022,Xu2021}.
A further physical constraint is that the input potential $V_i$ of a node is not given by a simple weighted sum of the output potentials $V_o$ from  the previous layer.  Instead, Kirchhoff's law requiring zero  net current at node $o$  introduces a normalization factor. The resulting node potential is therefore 
\begin{equation}
    V_o = \frac{\sum_i^N G_{o,i} V_i}{\sum_i^N G_{o,i}},
    \label{eq:physical}
\end{equation}
where $N$ input nodes are connected to the output node through conductances $G_{o,i}$, which play the same role as the weights $W_{ij}$. We refer to them as  conductances to emphasize the underlying physical interpretation.
The resulting potential can then be passed through an activation function to introduce  nonlinearity into the network.
In passive systems, the first law of thermodynamics further constrains  the conductances to be positive. Here, however, we  assume that we have access to active circuit elements that can result in an effective negative conductance, e.g.\ through operational amplifiers (op-amps), see Appendix \ref{sec:circuits}.
We note that machine learning without negative weights is in principle possible if one employs an activation function with both positive and negative slopes, such as \ $f_a(x) = 1/(1+(2x-1)^2)$. However, this approach no longer succeeded once the physically constrained node rule of \cref{eq:physical} was used as well (see Appendix \ref{sec:approx}).
\begin{figure}
    \centering   \includegraphics[width=\linewidth]{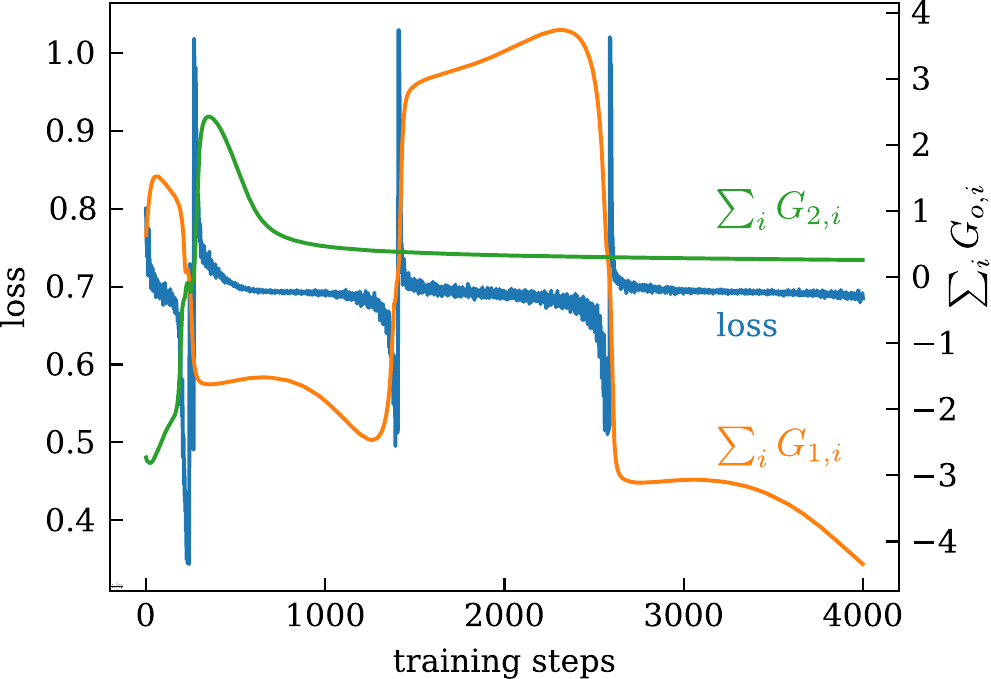}
    \caption{Loss in a trivial dense $2 \times 2$ network trained to classify points above and below a straight line. Sudden increases in the loss function occur whenever $\sum_iG_{o,i}$ changes sign.}
    \label{fig:losses}
\end{figure}

Allowing for negative weights can introduce an additional complication: the total conductance $G_o=\sum_i G_{o,i}$ can  change sign upon incremental updates of the individual $G_{o,i}$ during training. When this occurs, the effective node input undergoes a sign inversion, which can severely disrupt learning when standard machine-learning activation functions are used.
This effect already appears in extremely simple networks and datasets. To illustrate this, we consider  a fully connected network with two inputs and two outputs trained to classify whether a point $(x,y)$ lies above or below an affine linear function $f(x)=ax+b$. During training, the loss initially decreases smoothly as shown by the blue curve in \cref{fig:losses}, but then abruptly jumps back to values corresponding to an essentially untrained network, with nearly 100\% misclassification. These jumps occur precisely when $\sum_i G_{1,i}$ (orange curve) or $\sum_i G_{2,i}$ (green curve) at one of the two output nodes 
 changes sign, causing the input to node $o$ to switch from $V_o$ to $-V_o$, and consequently changing the output of node $o$ from $f_a(V_o)$ to $f_a(-V_o)$.

We identified two approaches to circumvent this issue, which we term \emph{normalized physical machine learning} (npML) and \emph{symmetrized physical machine learning} (spML).
In npML, we explicitly prevent $G_o$ from changing sign by adding a regularization term to the loss function. The modified loss function becomes
\begin{equation}
    L_m = L + \lambda \sum_o|G_o - 1|,
\end{equation}
where $L$ is the original loss function, $\lambda=0.1$ is the regularization parameter, and the sum runs over all nodes $o$  in the network.
This term stabilizes the normalization factor by biasing $G_o$ toward positive values close to unity. With this  regularization, the physical model shown in  \cref{fig:ml-setup} successfully learns  nontrivial datasets with performances comparable to conventional machine learning (see the npML and ML column  in \cref{tab:L2readout} for quantitative results).
If modifying the loss function is 
undesirable, a second strategy is to  symmetrize the activation function, which we refer to as spML. In this approach, we replace the standard activation function $f_a(x)=\sigma(5x)$  by  $f_a(x) = \sigma(5(|x|-0.5))$  for the hidden nodes.
This symmetrization makes the network insensitive to sign flips of the node input and yields performance comparable to the regularization-based approach. 
From a hardware perspective, spML may be particularly attractive because it avoids modifications of the loss function itself. This could enable the use of contrastive-learning approaches that operate without explicit backpropagation, thereby substantially reducing the energy cost of training.

\section{Conclusion}
In this work, we have shown that non-linearity, as well as  the ability to amplify and suppress signals, are essential ingredients for physical neural networks capable of performing general machine-learning tasks. We further clarified how passive linear networks can perform successful classification, even though they should not have the required expressivity. This is achieved through the use of powerful  nonlinear readout schemes applied at the output stage, typically implemented on an external computer. 
To evaluate the performance of a physical network, it is important to systematically compare the performance obtained with and without the proposed physical computing element, especially when unconventional readout schemes are used. In addition, care should be taken when using datasets that are already linearly or nearly linearly separable (such as Iris, Penguin, Wine, Rice, or Wholesale Customers) as benchmarks for physical learning, since a significant fraction of the performance may already arise from the readout function rather than from the physical network itself. 

If the underlying device is effectively linear, it cannot serve as a general-purpose computing platform for nonlinear tasks. In this sense, such architectures may be viewed as a form of \emph{inverse reservoir computing}: instead of a complex nonlinear dynamical system followed by a simple linear readout, they rely on a trivial linear “reservoir” combined with a complex nonlinear readout function that performs most of the computational work.

We conclude by noting several limitations of the present work. Although we demonstrate physically realistic networks that are capable of  learning, their implementation relies on active components, which partially reduces the  energy-efficiency advantages typically associated with  physical computing platforms. Additionally, standard machine-learning techniques such as batch normalization are not straightforward to apply  within the current architecture. Finally, on-device training may be challenging, since backpropagating information through the nonlinear elements of the system is not naturally supported and would likely require additional architectural or circuit-level mechanisms.

\section{Data availability}
All data can be generated with the source-code provided in~\cite{code}.
The raw data is also available in Ref.~\cite{data}.

\section{Code availability}
The code to perform the simulations is provided in~\cite{code}.

\section{Competing Interests}
There are no competing interests.

\section{Author Contributions}
N.C.X.S performed the simulations and analyzed the data. Both authors discussed the results and contributed to the manuscript.

\section{Acknowledgements}
Special thank to Edwin Bedolla for many helpful discussions about machine learning.
N.C.X.S and M.D.\ acknowledge funding
from the European Research Council (ERC) under the
European Union’s Horizon 2020 research and innovation
program (Grant agreement No. ERC-2019-ADG 884902,
SoftML) as well as the Marie Skłodowska-Curie Actions program (Grant agreement ID 101206265, BRAIN-CCC).
\bibliography{bib}

\appendix
\clearpage

\section{Contrastive learning rule}
\label{sec:contrastive}
For training the network using a  contrastive local update rule, two identical copies of the same network are required~\cite{Nachi}.
In the \emph{free} phase,  input voltages are applied to the input nodes and the resulting output voltages $V^f$ of the network are measured at the output nodes. To keep the notation concise, we write a single output voltage, although the formulation straightforwardly generalizes to multiple output voltages.
To train the network, the output of the free network is compared to the desired output $V^d$. A clamped target voltage is then constructed as a linear interpolation between the two,  $V^c=\eta V^d + (1-\eta)V^f$, where $\eta$ denotes the learning rate. This clamped voltage is applied to the output nodes of the second copy of the system, referred to as the \emph{clamped} network.
The voltage drop $\Delta V^f$ and $\Delta V^c$ across each edge are then measured in both the free and clamped configurations, respectively, and the corresponding resistance (or conductance) values are  updated using a purely local update rule based on their difference.

\begin{figure}[htbp]
  \centering
  \includegraphics[width=\linewidth]{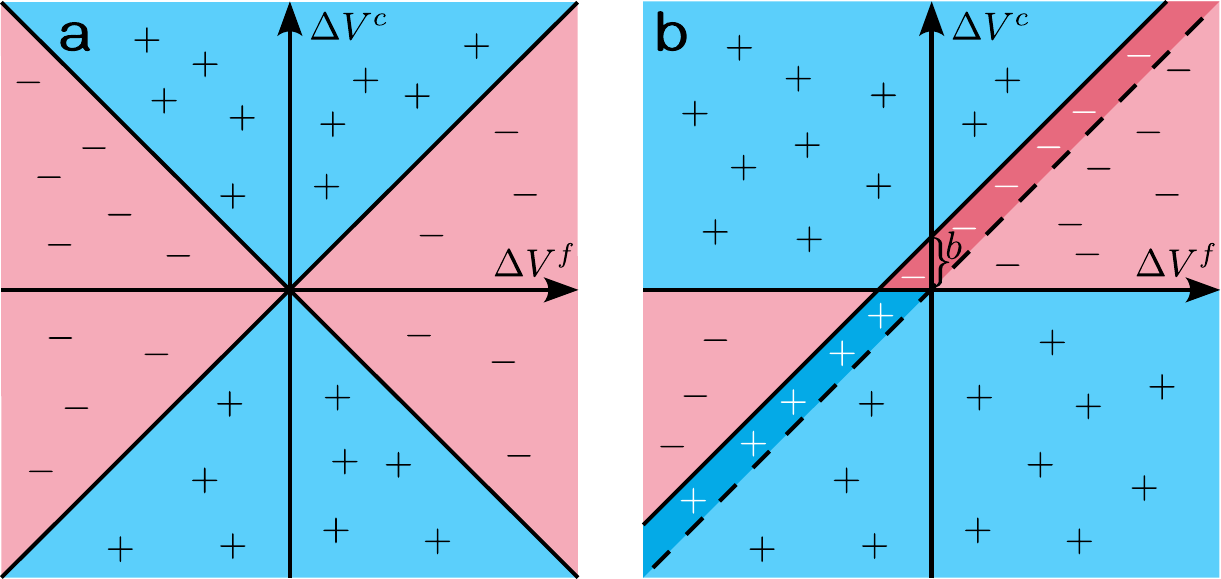}
  \caption{Comparison of the two update rules \cref{eq:main} (panel (a)) and the sign-corrected \cref{eq:XORwrong} (panel (b)) from  Ref.~\cite{Nachi} shown as a function of the clamped voltage drop $\Delta V^c$, and free voltage drop $\Delta V^f$ across a resistor. Blue regions marked with plus signs indicate an increase in resistance, while pink regions marked with minus signs indicate a decrease in resistance. The shift of the decision boundary in panel (b) illustrates the effect of the bias $b$. This bias skews the update regions, making resistance increases more likely in some parts of the $(\Delta V^f,\Delta V^c)$ space  (darker blue), while inducing the opposite  in other regions (dark pink).
  }
  \label{fig:stupid-rule}
\end{figure}

The authors of Ref.~\cite{Nachi} present  two update rules in their work, which they claim are equivalent. However, these two formulations are not consistent, even after correcting  a sign error in their derivation.
The update rule given in their main text reads:
\begin{equation}
  \label{eq:main}
  \Delta R =
  \begin{cases}
    \phantom{-}  \delta R & \text{if } |\Delta V^f| < |\Delta V^c|\\
    - \delta R &\text{else}\\
  \end{cases},
\end{equation}
which is stated to be equivalent to
\begin{equation}
  \label{eq:XORwrong}
  \Delta R =
  \begin{cases}
    \phantom{-}  \delta R &\text{if } \mathrm{xor}(\Delta V^f < \Delta V^c, 0 < \Delta V^c)\\
    - \delta R &\text{else}\\
  \end{cases},
\end{equation}
in the case where the voltage drops have the same sign, i.e.\ $\sgn(\Delta V^f) = \sgn(\Delta V^c)$. However, this equivalence does not hold as there is a sign error. The logical condition must be inverted for the two expressions to agree.
In general, the voltage drops in the free and clamped states do not necessarily have the same sign. In the experiments of Ref.~\cite{Nachi}, the authors report that the signs coincide in more than 90\% of the cases. However, this observation arises primarily in a regime where the system is already close to its maximal performance and the output is effectively dominated by the $L_2$-readout. In more general settings, this sign agreement is not guaranteed. When the sign condition fails, the update rule no longer behaves as intended, and the network can be driven in an incorrect direction in parameter space, thereby hindering or even reversing learning progress.
Additionally, in Fig.~7(e) of Ref.~\cite{Nachi}, the two curves do not intersect at zero voltage difference but instead at approximately $0.003\,V$, indicating the presence of a small bias term $b$ in the update rule originating from imperfections in the $\mathrm{xor}$-gates. This effectively modifies the condition in  \cref{eq:XORwrong} to $\mathrm{xor}(\Delta V^f +b < \Delta V^c, 0 < \Delta V^c)$. 
The authors argue that this bias enforces a systematic increase in the resistances. However, this behavior only holds under the condition $\Delta V^c>0$. For $\Delta V^c<0$, the bias instead acts in the opposite direction, effectively reversing the intended update and introducing a sign-dependent asymmetry in the learning dynamics. See \cref{fig:stupid-rule} for further details.

We introduce a new contrastive local learning rule, which we term \emph{Ignore Free}. This rule corresponds to the limit $b\rightarrow\infty$. In this limit, the condition in \cref{fig:stupid-rule} reduces to $\mathrm{xor}(\mathrm{false}, 0<\Delta V^c)$, which simplifies the update rule to
\begin{equation}
\label{eq:ignore}
\Delta R =
\begin{cases}
\phantom{-} \delta R & \text{if } 0 < \Delta V^c \\
- \delta R & \text{otherwise.}
\end{cases}
\end{equation}
Although we refer to this rule as \emph{Ignore Free}, it does not fully discard information from the free phase. Instead, information about the free state is implicitly encoded in the clamped voltage drops, since the clamped input voltages are given by $V^c = V^f + \eta (V^d-V^f)$.
In Appendix \ref{sec:reorder}, we compare the three learning rules \cref{eq:main,eq:XORwrong,eq:ignore} and study how their performance depends on the subset of Iris dataset samples used during training.

\section{Stability of different local learning rules}
\label{sec:reorder}
\begin{figure}[htbp]
  \centering
  \includegraphics[width=\linewidth]{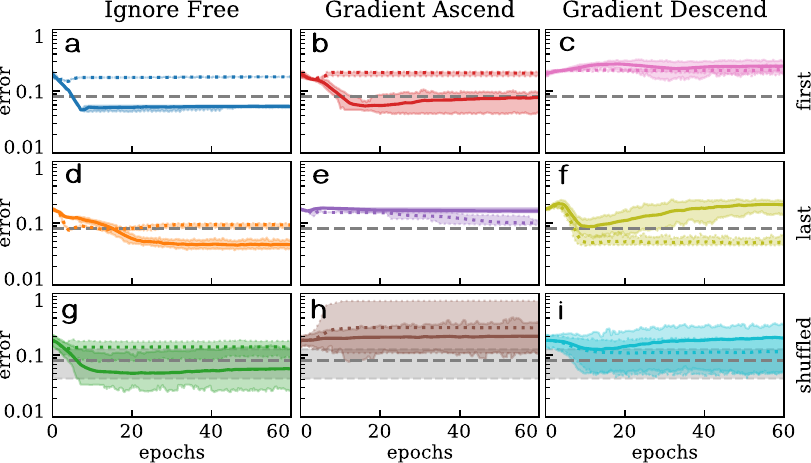}
  \caption{Classification error for different learning rules (columns) and different ways of splitting the dataset into training and test sets (rows). In the first row (a-c), the first ten samples of each flower class are used for training. In the second row (d-f), the last ten samples of each class  are used for training. In the third row (g-i), ten samples per class are selected randomly for training. Columns correspond to different learning rules: The first column uses the update rule in \cref{eq:ignore}, which  ignores the free state; the second column uses  gradient ascent based on  \cref{eq:XORwrong}; and the third column shows standard gradient descent.   Dotted lines indicate simulations in which the desired voltages are kept fixed throughout training, whereas solid lines correspond to the update scheme of Ref.~\cite{Nachi}, in which the desired voltages are updated after each training epoch. Shaded regions indicate the 95\% confidence interval obtained from 100 independent simulations per configuration. The dashed line denotes the baseline performance of the readout rule without any network. When the error exceeds this baseline, the network degrades performance rather than improving it. }
  \label{fig:iris}
\end{figure}

In this section, we take a closer look at the three learning rules introduced in \cref{eq:main,eq:XORwrong,eq:ignore} and investigate their robustness with respect to how  the Iris dataset is partitioned into  training and testing samples. To assess whether the network genuinely contributes to classification, we additionally compare its performance against the baseline case in which no network is used and the post-processing classifier is applied directly to the input data.
We also examine whether updating the desired output voltages during training~\cite{Nachi}  improves performance. The combined results are summarized in \cref{fig:iris}.
We find that the rule that largely ignores  the free state consistently yields the best performance. The methods proposed in Ref.~\cite{Nachi} only succeed under rather specific conditions. In particular, the gradient-ascent (!) approach performs reasonably well when the first ten samples of each class are used for  training and the desired voltages are updated after every epoch. However, after $\approx 40$ epochs, the classification accuracy drops below the baseline obtained without any network at all.
This degradation occurs because the repeated adjustments of the desired voltages drives the network resistances toward extreme values, creating a negative feedback loop that continuously reduces the effective resolution of the network. We  note that gradient ascent is known to systematically remove, rather than learn, information in neural networks~\cite{Melamed2025}. 
We therefore question whether the  network is learning in this regime. 
Instead, we argue that the apparent success of gradient ascent is largely a peculiarity  of the Iris dataset together with the particular subset of  samples chosen for training. Indeed, the same procedure fails when either  the last ten samples or randomly selected  samples from each class are used for training.

A similar argument applies to the gradient-descent approach, which only consistently outperforms the baseline without a network when the last ten  samples of each class are used for training and the desired voltages are kept fixed throughout training. Based on these observations, we suspect that the authors of Ref.~\cite{Nachi} effectively operated in a regime corresponding to gradient descent and  reported the training performance only  up to the point of minimal error. Our results indicate that, if training is continued for additional epochs, the network gradually loses the information it had previously acquired, leading to a deterioration in classification performance.
In contrast, we find that the networks learn consistently when the \emph{Ignore Free} update rule is used and the desired voltages are adapted during training. Only this approach remains robust when random subsets of samples from each class are chosen for  training, indicating that it captures genuinely reproducible learning behavior rather than exploiting particularities of a specific train/test split.

\section{Effects of system size}
\label{sec:systemsize}
\begin{figure}[htbp]
  \centering
  \includegraphics[width=\linewidth]{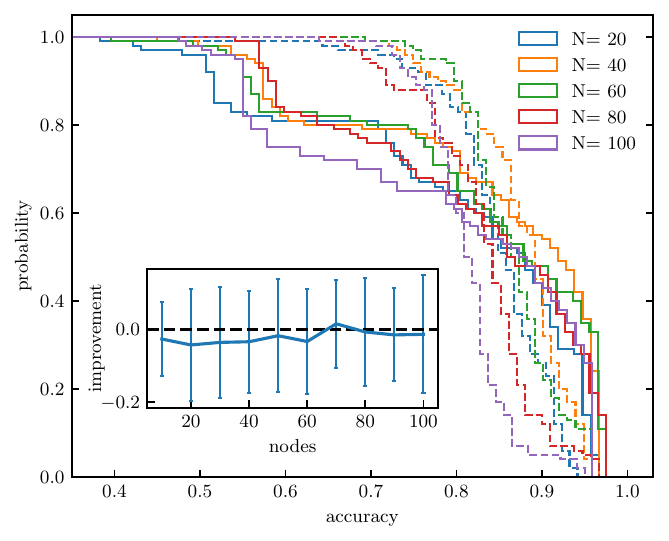}
  \caption{Cumulative histograms of the classification accuracy for  networks of different sizes, where the node count $N$ is indicated in the legend. For a network with $N$ nodes, the  height of the corresponding curve at accuracy $a$  gives the probability that a randomly initialized network of that size achieves an accuracy less than or equal to $a$. The dashed curves show  the accuracies before training. Training broadens the distribution of accuracies, indicating that some networks  improve while others deteriorate, but it does not produce a systematic increase in the mean performance. This is further illustrated in the inset, which shows the average accuracy change for 100 networks of each size; error bars denote the standard deviation. On average, training reduces the network performance by a few percentage points. 
  }
  \label{fig:size}
\end{figure}

It is tempting to assume that the poor  classification accuracy is due to the small size of the networks, and that larger networks would perform better. 
To test whether increasing network size improves performance on the Iris dataset, we systematically generate random connected Watt-Strogatz graphs~\cite{Watts1998} with sizes  ranging from 10 to 100 nodes. We set the expected node degree to $4$ and the rewiring probability to $0.3$. Input and output nodes are selected randomly for each graph. For each size, we generate 100 independent graphs and train them using 10 randomly selected samples from each  flower class over 20 epochs. We plot the initial and final classification accuracies in \cref{fig:size}.
We find that the learning performance is largely independent of network size, with no clear trend, indicating improved performance for larger networks. Moreover, the average performance after training is not higher than before training. In fact, networks tend to perform slightly worse after training, as  shown in the inset of \cref{fig:size}. 
It is also unsurprising  that larger networks do not outperform smaller networks. Increasing the size of a  linear network does not increase its expressive power; it only increases  the resolution with which the underlying linear map can be represented.  As a result, the entire network can still be represented by an effective  $4\times 3$ linear mapping, regardless of its internal size.

We  note that the best-performing configuration for classifying the Iris dataset with the $L_2$-classifier is a minimal two-node network consisting of a single input and a single output, where the petal width is directly fed into the $L_2$-classifier. This setup achieves an accuracy of $\approx 95\%$. Since the system reduces to a single resistor, the specific resistance value is irrelevant: the output voltage is simply equal to the input voltage.  
Depending on the training/testing data split, this two-node configuration yields an accuracy of $95.8\%$ with a standard deviation of $1.1\%$ over 100 random splittings of the dataset.

\section{Relaxing constraints}
\label{sec:approx}
\begin{figure}[htbp]
  \centering
  \includegraphics[width=\linewidth]{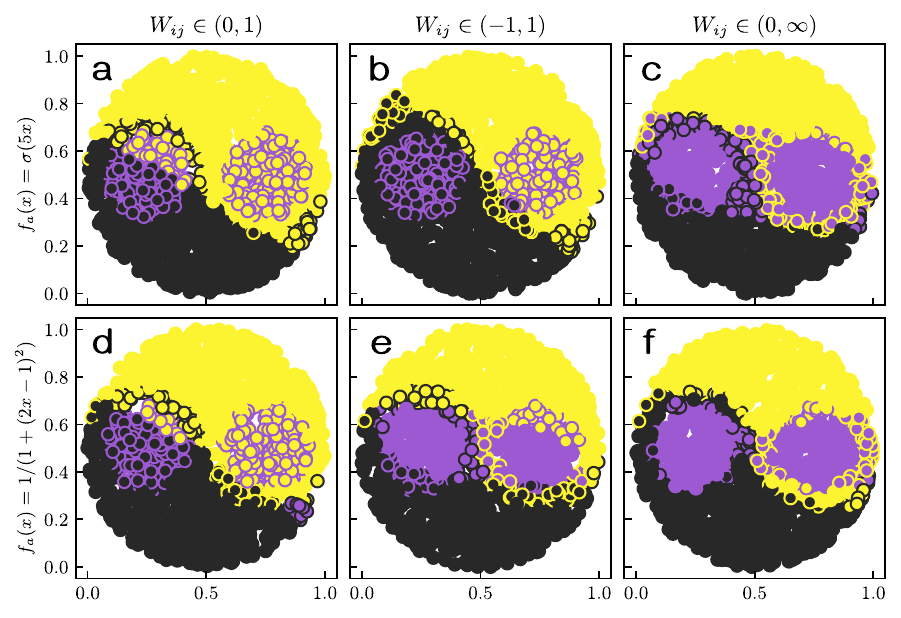}
  \caption{
  Classification of the Yin-Yang dataset under restricted weight conditions. The top row (a-c) uses a  monotonically increasing activation function, while the bottom row (d-f) employs a non-monotonic activation function. In first column, weights are restricted to the interval  $(0,1)$, in the second column to $(-1,1)$, and in the third column to $(0,\infty)$. Outer circles 
  represent the true class labels, while inner circles indicate the predicted labels. 
  Only panels (e,f) qualitatively capture the Yin-Yang structure. In all other cases, the classification performance is strongly degraded. In particular, panel (f), where both  amplification and suppression are  possible, reproduces the Yin-Yang structure most accurately. }
  \label{fig:restricted_yinyang}
\end{figure}

In the main text, we discussed the behavior of  physically realistic networks under strict physical constraints. When active components are introduced, these constraints can be relaxed. In this section, we therefore explore the effect of less restrictive constraints on both the weights and activation functions.

To impose bounds on the weights, we introduce  an additional activation function acting on  the weights, i.e.\ $w_{ij} = f_w(\hat{w}_{ij})$, where $\hat{w}_{ij}\in \mathbb{R}$ denotes the unconstrained internal weights of the network optimized during training, and $f_w$ maps $\mathbb{R}$ to a specified range. We consider $f_w\in \{\tanh, \sigma, \exp\}$,  which constrain the weights  $w_{ij}$ to the intervals $(-1,1)$, $(0,1)$ and $(0,\infty)$, respectively. 
We choose $\tanh$ to emulate systems with potentially negative weights but without  amplification, $\sigma$ to model systems with strictly positive bounded weights and no amplification, and $\exp$ to represent systems with  positive weights that allow amplification.
We note that, although individual weights are  bounded (e.g. in (0,1) or (-1,1)), the inputs to the activation function are not bounded. 
Since these contributions are summed at each node, the resulting node input can exceed the range of a single weight. This is consistent with the physical interpretation in terms of potentials and currents: while each edge current is limited by the conductance range, currents arriving from multiple edges can add up, and the node can receive a total input signal that exceeds the maximum conductance times the output potential of the previous node. 
In addition to restricting the weights, we introduce a non-monotonic activation function
\begin{equation}
    f_a(x)=\frac{1}{1+(2x-1)^2}
    \label{eq:decr}
\end{equation}
 to half of our numerical experiments. This  function remains bounded in (0,1), consistent with systems  constrained to a finite voltage range. While remaining bounded, this function exhibits both positive and negative slopes. Since its derivative changes sign, it enables both amplification and suppression of input signals, which are essential for non-trivial information processing.  This is not possible with monotonically increasing activation functions when all weights are positive.
Without such non-monotonicity, the entire network becomes monotonic, meaning that inputs with larger magnitude will always produce outputs with larger magnitude. 
This would severely restrict the space of  functions  in which the network can approximate  the  target function $f_d$.
We can observe this effect in \cref{tab:resticted_weights} and \cref{fig:restricted_yinyang}, where we compare networks under the aforementioned weight constraints using both monotonic and non-monotonic activation functions across various datasets.
Functions with a shape similar to \cref{eq:decr} should be implementable in hardware using a bandpass response.

\begin{table}[htbp]
  \centering
  \begin{tabular}{l|cccccc|c}
	  Dataset & $^{1}_{0} r^{+}$& $^{1}_{0}r^{\pm}$ & $^{+1}_{-1}r^+$ & $^{+1}_{-1}r^{\pm}$ & $^\infty_{0}r^+$ & $^\infty_0r^{\pm}$ & ML\\
 \hline
Iris     & $39\%$ &  $88\%$ &  $71\%$ &  $96\%$ &  $61\%$ &  $95\%$ & $93\%$\\
Penguins & $45\%$ &  $79\%$ &  $80\%$ &  $97\%$ &  $73\%$ &  $95\%$ & $98\%$\\
Wine     & $39\%$ &  $68\%$ &  $65\%$ &  $97\%$ &  $78\%$ &  $97\%$ & $96\%$\\
Rice     & $57\%$ &  $92\%$ &  $92\%$ &  $93\%$ &  $92\%$ &  $92\%$ & $93\%$\\
Customer & $68\%$ &  $85\%$ &  $80\%$ &  $90\%$ &  $90\%$ &  $90\%$ & $90\%$\\
Magic    & $65\%$ &  $81\%$ &  $78\%$ &  $85\%$ &  $83\%$ &  $86\%$ & $86\%$\\
\hline
Yin Yang & $55\%$ &  $65\%$ &  $65\%$ &  $76\%$ &  $74\%$ &  $80\%$ & $89\%$\\
Circle   & $53\%$ &  $77\%$ &  $60\%$ &  $96\%$ &  $67\%$ &  $98\%$ & $89\%$\\

  \end{tabular}
	\caption{
    Machine learning accuracy under different restrictions on  weights and the derivative of the activation function, compared to classical machine learning. All networks follow the topology shown in \cref{fig:ml-setup}.
    Mean accuracy over 25 independent training runs  is reported. Column labels $^a_br^c$ indicate the imposed constraints. The left sub/superscripts specify the range of allowed  weights, i.e.\ $w_{ij}\in (a,b)$. The right superscript indicates constraints on the derivative of the activation function: $+$ indicates a monotonic  activation function, while $\pm$ indicates a non-monotonic  activation function. Restricting either the weights or the  activation function can significantly reduce the accuracy of the network. 
    In particular, the results indicate that effective learning requires both signal amplification and suppression; only columns 4 and 6 achieve consistently high performance. See \cref{fig:restricted_yinyang} for the corresponding classification of the Yin–Yang dataset of the best performer under each network constraint.}

  \label{tab:resticted_weights}
\end{table}

\section{Circuit Realization of the Proposed Physical Network}
\label{sec:circuits}

\begin{figure}[htbp]
    \centering
    \includegraphics[width=0.9\linewidth]{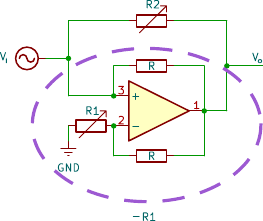}
    \caption{Possible circuit realization of an adjustable (negative) resistance. The circuit enclosed by the dashed line exhibits an apparent negative resistance of $-R_1$. Variable resistors $R_1$ and $R_2$ control the magnitudes of the negative and positive resistances, respectively.  The absolute resistance value of the resistors labeled  $R$ is unimportant, provided that all resistors $R$ have identical resistance values. An input voltage $V_i$ is applied in parallel to the positive and negative resistors, producing the output voltage $V_o$.}
    \label{fig:circuit}
\end{figure}

Apparent negative resistances can be  realized using operational amplifiers (op-amps). We propose a parallel connection of a conventional tunable resistor $R_2$ and a tunable apparent negative resistance $-R_1$, as shown in \cref{fig:circuit}. When an input voltage $V_i$ is applied to the circuit, it produces an output voltage $V_o = \frac{-R_1 R_2}{R_2 - R_1}V_i$. 

\end{document}